\documentclass[12pt]{article}
\title{\huge Surface coupling effect on wetting and layering transitions} 
\author{\bf  
 L. Bahmad, A. Benyoussef and H. Ez-Zahraouy%
\thanks{Corresponding author: ezahamid@fsr.ac.ma}
\\
 Laboratoire de Magn\'{e}tisme et de la Physique
 des Hautes Energies
\\
Universit\'{e} Mohammed V, Facult\'{e} des Sciences, Avenue Ibn Batouta,  B.P. 1014
\\Rabat, Morocco
}
\date{ }
\begin{document}
\maketitle 
\begin{abstract}
The effect of the surface coupling $J_s$ on the dependency of the layering transition temperature $T_L$ as a function of the thickness $N$, of a spin$-1/2$ Ising film, is studied using the mean field theory. It is found that for $J_s$ greater than a critical value ($J_{sc}=1.30$), the layering transition temperature decreases when the film thickness $N$ increases for any values of the surface magnetic field $H_s$. While, for $J_s < J_{sc}$, the behaviour of the layering transition temperature $T_L$, as a function of $N$, depends strongly on the  values of $H_s$. Indeed, we show the existence of three distinct behaviours of $T_L$, as a function of the film thickness $N$, separated by two critical surface magnetic fields $H_{sc1}$ and $H_{sc2}$, namely: $(i)$ for $H_s < H_{sc1}$, $T_L$ increases with $N$; $(ii)$ for $H_{sc1} < H_s < H_{sc2}$, $T_L$ increases for small values of $N$, and decreases for large value ones; (iii) while for $H_s > H_{sc2}$, $T_L$ decreases with increasing the film thickness. Furthermore, depending on the values of $J_s$, the wetting temperature $T_w (T_w=T_L$ when $N \rightarrow \infty$ for a given material), can be greater or smaller than the layering transition temperature of a film of thickness $N$ of the same material.
\end{abstract}
\noindent



\section{Introduction}
\mbox{  } The wetting and layering transitions of magnetic Ising systems have been studied by several authors. 
A simple lattice gas model with layering transitions and critical points has
been introduced and studied in the mean field approximation by de Oliveira
and Griffiths [1], Pandit {\it et al.} [2], Nightingale {\it et al.} [3] and Ebner {\it et al.} [4-7]. A variation of phase diagrams with the strength of the substrate potential in lattice gas model for multi-layer adsorption has been studied, using Monte Carlo simulations, by Patrykiejew {\it et al.} [8], and Binder and Landau [9].
One type of transitions is the layering transitions, in which the thickness of a solid film increases discontinuously by one layer as the pressure is increased. Such transitions have been observed in a variety of systems including for example $^{4}He$ [10,11] and ethylene [12,13] adsorbed on graphite. \\
Ebner [14] carried out Monte Carlo simulations of such a lattice gas model. Huse [15] applied renormalization group technique to this model. It allowed the study of the effects on an atomic scale or order disorder transitions in the adsorbed layers, which may have considerable influence on the layering transitions and tracing back macroscopic phenomena on inter-atomic potentials. 
Benyoussef and Ez-Zahraouy have studied the layering transitions of Ising model thin films using a real space renormalization group [16], and transfer matrix methods [17].
As for the film system, which is finite in one direction, it has been established that its magnetic properties can differ greatly from those of the corresponding bulk [18-22]. Experimental results [23] showed that the critical temperature of a vanadium film depends on the film thickness and its critical behaviour is like that of the two-dimensional system rather than that of the three-dimensional bulk.  
Although the great number of theoretical works made in the field of wetting phenomena, especially the effect of the nature of the substrate potential on wetting and layering transitions [1,2,16,17,24,25], the relation between the surface coupling and the wetting temperature is not yet sufficiently investigated. However, Hong [26] has studied the effect of the surface couplings on the behaviour of the critical temperature film as a function of its thickness. \\
\mbox{  }Our aim in this paper is to study the effect of the surface coupling (different than the bulk one) on the behaviour of the layering transition and the wetting transition temperatures, as a function of the film thickness of a spin$-1/2$ Ising film, using the mean field theory. \\
However, depending on the value of the surface coupling $J_s$, the layering transition temperature $T_L$ can increase or decrease with the film thickness and reaches the wetting temperature $T_w (T_w=T_L$ when $N \rightarrow \infty$) for a sufficiently thick film.
Furthermore, it is found that there exists a critical value $H_{sc}$ of the surface magnetic field below which the wetting temperature is independent on the surface coupling $J_s$, while for $H_s$ greater than $H_{sc}$ the wetting temperature depends on the $J_s$ values. Such results can be useful for experiments.
This paper is organised as follows. Section 2 describes the model
and the method. In section 3 we present results and discussions.
\\
\section{Model and method}
We consider a film formed with N coupled ferromagnetic square layers $k=1,...,N$ ($k=1$ is considered as a surface) infinite in the $x$ and $y$ directions, in an external and surface magnetic fields, as it is illustrated by Fig. $1$. \\
It is well known that the spin$-1/2$ variable values can be written as: $S_i=(\hbar/2)\sigma_i$ where $\sigma_i =\pm 1$. 
For simplicity we assume, hereafter, that $\hbar/2=1$.  Hence, the Hamiltonian governing the system is given by:
\begin{equation}  
{\cal H}=-\sum_{<i,j>}J_{i,j}\sigma_{i}\sigma_{j}-\sum_{i}H_{i}\sigma_{i}
\end{equation}
The first sum runs over all nearest-neighbour sites. In several previous works [16,17,24,25], the case of a constant coupling $J_{i,j}=J_b=$constant has been studied. Here, we consider the model where the constant couplings are given by:
\begin{equation}   
J_{i,j}=\left\{ \begin{array}{ll}
	      J_{s} & \mbox{for $(i,j)$ $\epsilon$ surface}  \\
	      J_{b}       & \mbox{elsewhere} \\	     
	      \end{array}
	\right.
\end{equation}   
In the last sum of the Hamiltonian $(1)$, the total magnetic field $H_{i}$ is applied on each site $"i"$, and distributed according to: \\ 
\begin{equation}   
H_{i}=\left\{ \begin{array}{ll}
	     H+H_{s} & \mbox{for $i$ $\epsilon$ surface}  \\
	     H+H_{s}/k^\alpha    & \mbox{elsewhere} \\	     
	      \end{array}
	\right.
\end{equation} 
with $\alpha$ being a positive constant. In all the following, we will be limited to a constant surface magnetic field $H_{s}$ applied only on the surface $(k=1)$. This is the case for $\alpha \rightarrow \infty$. \\
The results of the model for a finite $\alpha$ value exhibits similar topologies of the phase diagrams as it was outlined by Binder {\it et al.} [27,28] and in some of our earlier works [16,17,29]. \\
Using the mean field theory, we introduce the effective Hamitonian
$H_0=\sum_ik_i\sigma_i+\sum_iH_i\sigma_i$,
with $k_i=\sum_j J_{ij}\sigma_j$, the sum runs over the nearest-neighbour sites $'j'$ of the site $'i'$. 
The magnetization of each site $i$ is given by:
$m_i=\frac{Tr \sigma_i exp(-\beta H_0)}{Tr exp(-\beta H_0)}$,
where $Tr$ means the trace performed over all spin configurations.
Since the system is invariant by translation  in both $x$ and $y$ directions, $m_i=m_k$
for each site 'i' of the layer k. Hence, the magnetisation of a layer $k$ can be written as: 
\begin{equation}   
\left\{ \begin{array}{ll}
m_{1}=\tanh(\beta(4J_sm_{1}+J_bm_{2}+H_1)) & \mbox{for k=1 } \\
m_{k}=\tanh(\beta(4J_bm_{k}+J_bm_{k-1}+J_bm_{k+1}+H_k)) & \mbox{for  k=2,...,N-1} \\
m_{N}=\tanh(\beta(4J_bm_{N}+J_bm_{N-1}+H_N)) & \mbox{for k=N} \\ 
\end{array}
	\right.
\end{equation}
With the free boundary conditions: $m_{0}=m_{N+1}=0$. \\
The total free energy of the system can be written as follows 
\begin{equation}
  F=\sum_{k=1}^{N} F_{k}, 
\end{equation}
where $F_{k}$ stands for:
\begin{equation}   
\left\{ \begin{array}{ll}
F_{1}=-\frac{1}{\beta}Log(2cosh(\beta(\lambda_1+H_1)))+\frac{1}{2}m_1\lambda_1 & \mbox{for k=1 }  \\
F_{k}=-\frac{1}{\beta}Log(2cosh(\beta(\lambda_k+H_k)))+\frac{1}{2}m_k\lambda_k & \mbox{for k=2,...,N}\\     
\end{array}
	\right.
\end{equation}
where 
\begin{equation}   
\left\{ \begin{array}{ll}
\lambda_1=4J_sm_{1}+J_bm_{2} & \mbox{for k=1} \\
\lambda_k=4J_bm_{k}+J_bm_{k-1}+J_bm_{k+1} & \mbox{for  k=2,...,N} \\ 
\end{array}
	\right.
\end{equation}
and $\beta=1/(k_{B}T)$ with 
$T$ being the absolute temperature and $k_{B}$ the Boltzmann constant.

\section{Results and discussion}
 Reduced values of the parameters $T$, $T_L$, $T_w$, $H$, $H_s$ and $J_s$ are investigated in this work. However, for simplicity the syntax "reduced" will be cancelled despite the fact that we use the notations $X/J_{b} (X=T, T_{L}, T_{w}, H, H_{s}, J_{s})$ and give numerical values of these reduced parameters.  \\
The notation $1^{k}O^{N-k}$ will be used for a configuration where the first top $k$ layers from the surface, are in the state 'up' and the remaining $N-k$ bottom layers are in the state 'down'. In particular, $1^N$ (rep. $O^N$) will denote a system where all the layers are with positive magnetisation (rep. negative magnetisation). There exists $N+1$ possible configurations for a film formed with $N$ layers. However, the transition of the layer "$k$" is characterised by the change of the sign of its magnetisation while the magnetisations of the remaining $N-k$ layers of the film keep their initial sign i.e.: $1^{k-1}O^{N-k+1} \leftrightarrow 1^{k}O^{N-k}$. Hence, for $k=1$ the transition is called the surface transition.\\
The ground state phase diagram $(T=0)$ do not depend on the surface coupling constant $J_s$. Hence, in the following we will give results and phase diagrams for $T \ne 0$.
We solve numerically the equations $(4)$ and $(6)$  in order to establish the phase diagrams of the system. 
Indeed, Fig. $2$ shows that the layering transition temperature $T_{L} /J_b$,
defined as the temperature above which the layers $(k\ne1)$ of the film  begin to transit layer-by-layer, which is usually greater than the "surface transition" temperature. Except the special case of a film formed with two layers, where the surface transition temperature is close to the layering transition temperature $T_{L} /J_b$.
It is understood that the wetting temperature $T_w /J_b$ coincides with the limit of $T_L/J_b$ for a sufficiently thick film $(N \rightarrow \infty)$. \\
A numerical study of the layering transition temperature as a function of the film thickens is illustrated by Figs. $3a$ and $3b$ for two surface magnetic field values and selected values of the surface coupling constant. 
It is shown that for a smaller value of the surface magnetic field, the wetting temperature keeps a constant value ($T_w /J_b \approx 5.1$ for $H_s /J_b=0.1$) at any surface coupling constant values, as it is shown in Fig. $3a$. 
For a higher value of the surface magnetic field (see Fig. $3b$), the scenario is inverted and the layering transition temperature $T_L /J_b$ (as well as the wetting temperature) depends strongly on the surface coupling values. Indeed, for a fixed value of the surface magnetic field $H_s /J_b=1.0$, some wetting temperature values are: $T_w /J_b \approx 5.5$ for a small value of the surface coupling constant $J_s /J_b=0.01$ and $T_w /J_b \approx 2.4$ for a higher value $J_s /J_b=1.8$, as it is summarized in Fig. $3b$. \\ 
Nevertheless, there exists three classes of the layering transition temperature behaviours, namely: 
The behaviour $(i)$ is a situation where the layering transition temperature $T_L/J_b$ increases with the film thickness and stabilises at a certain fixed value. This behaviour is always seen for small values of the surface coupling and any surface magnetic field value. The behaviour $(ii)$, where $T_L/J_b$ increases until a certain film thickness above which it decreases exhibiting a maximum. This situation is found for medium values of both the surface coupling and the surface magnetic field. Finally the behaviour $(iii)$, where $T_L/J_b$  decreases continuously and stabilises at a value close to $T_w/J_b$. This is generally the case for higher values of surface coupling. \\
It is worth to note that the wetting temperature $T_w/J_b$ (which coincides with $T_L/J_b$ for a sufficiently thick film) decreases as $H_s/J_b$ increases at a fixed surface coupling constant $J_s/J_b$. This is because the surface region becomes magnetically harder than the deeper layers, and as the film is thicker, the strength of the internal magnetic field decreases leading to a decrease of $T_w/J_b$. \\
We found two critical surface coupling constants $J_{sc1}/J_b$ and $J_{sc2}/J_b$. $J_{sc1}/J_b$ separates the behaviours $(i)$ and $(ii)$, while $J_{sc2}/J_b$ separates the behaviours $(ii)$ and $(iii)$. The dependency of these critical parameters, $J_{sc1}/J_b$ and $J_{sc2}/J_b$, as a function of surface magnetic field is plotted in Fig. $4$. For small values of the surface magnetic field $H_s/J_b$, the constants $J_{sc1}/J_b$ and $J_{sc2}/J_b$ become similar and are close to the value $1.30$, and vanish as the surface magnetic field $H_s/J_b$ increases. \\
The results of the Figs. 3(a,b) are summarised in Fig. $4$, since for small values of surface magnetic field $H_s/J_b$ the three behaviours of $T_L/J_b$, namely $(i)$, $(ii)$ and $(iii)$ are found in this figure, when increasing the surface coupling $J_s/J_b$ values.  \\
The two critical surface coupling constants $J_{sc1}/J_b$ and $J_{sc2}/J_b$ correspond, respectively, to two critical surface magnetic fields $H_{sc1}/J_b$ and $H_{sc2}/J_b$ separating the three regimes of the $T_L/J_b$ behaviours. 
For sufficiently large values of $J_s/J_b$ ($J_s/J_b \ge 1.30$), $T_L/J_b$ decreases when increasing the film thickness $N$ for any values of $H_s/J_b$. While, for $J_s/J_b < 1.3$, the layering transition temperature $T_L/J_b$ exhibits the three behaviours: (i), (ii) and (iii)  depending on the $H_s/J_b$ value. In particular for $H_s/J_b \ge H_{sc2}/J_b$, $T_L/J_b$ is always decreasing when the film thickness increases, see Figs. 3a, 3b.
It is also interesting to examine the surface coupling effect on the layering transition temperature, for a fixed surface magnetic field value. As it is illustrated in Fig. $5$, for a film size $N=5$ layers and several surface magnetic field values, the layering transition temperature increases for small values of $H_s/J_b$ and decreases when $H_s/J_b$ reaches higher values. This figure shows also that the layering transition temperature decreases when increasing the surface magnetic field $H_s/J_b$, for a fixed value of the surface coupling $J_s/J_b$. This result was already outlined in earlier works e.g. Pandit {\it et al.} [2] and in some of our recent works [24,25]. 

\section{Conclusion}      
\mbox{~~~ }In conclusion, we have studied the effect of the surface coupling $J_s$ on the wetting and layering transition temperatures as a function of the thickness $N$, of a spin$-1/2$ Ising film, using mean field theory. 
We showed the existence of a critical value ($J_{sc}=1.30$) of the surface coupling constant $J_s$, above which the layering transition temperature $T_L$ decreases when the film thickness increases for any values of the surface magnetic field $H_s$. While, for $J_s < J_{sc}$, there exists three distinct behaviours, namely:
$(i)$ for $H_s < H_{sc1}$, $T_L$ increases with $N$; $(ii)$ for $H_{sc1} < H_s < H_{sc2}$, $T_L$ increases for thin films, and decreases for thick ones; (iii) while for $H_s > H_{sc2}$, $T_L$ decreases with increasing the film thickness. Furthermore, for $J_s> J_{sc}$ the wetting temperature $T_w (T_w=T_L$ when $N \rightarrow \infty$), is independent on $J_s$. While for $J_s < J_{sc}$ $T_w$ depends on the value of $J_s$. Such result may be useful for experiments. \\
\noindent{\bf References}
\begin{enumerate}

\item[{[1]}] M. J. de Oliveira  and R. B. Griffiths , Surf. Sci. {\bf 71}, 687 (1978).
\item[{[2]}] R. Pandit, M. Schick and M. Wortis, Phys. Rev. B {\bf 26}, 8115 (1982); \\
$\mbox{}$ R. Pandit and M. Wortis, Phys. Rev. B {\bf 25}, 3226  (1982).
\item[{[3]}] M. P. Nightingale, W. F. Saam and M. Schick, Phys. Rev. B {\bf 30},3830 (1984).
\item[{[4]}] C. Ebner, C. Rottman and M. Wortis, Phys. Rev. B {\bf 28},4186  (1983).  
\item[{[5]}] C. Ebner and W. F. Saam, Phys. Rev. Lett. {\bf 58},587 (1987).
\item[{[6]}] C. Ebner and W. F. Saam, Phys. Rev. B {\bf 35},1822 (1987).
\item[{[7]}] C. Ebner, W. F. Saam and A. K. Sen, Phys. Rev. B {\bf 32},1558 (1987).
\item[{[8]}] A. Patrykiejew A., D. P. Landau and K. Binder, Surf. Sci. {\bf 238}, 317 (1990).
\item[{[9]}] K. Binder in {\it Phase Transitions and Critical Phenomena}, edited by C. Domb and J. L. Lebowits (Academic, New York, 1983), Vol. 8.
\item[{[10]}] S. Ramesh and J. D. Maynard, Phys. Rev. Lett. {\bf 49},47 (1982).
\item[{[11]}] S. Ramesh, Q. Zhang, G. Torso and J. D. Maynard, Phys. Rev. Lett. {\bf 52},2375 (1984).
\item[{[12]}] M. Sutton, S. G. J. Mochrie  and R. J. Birgeneou, Phys. Rev. Lett. {\bf 51},407 (1983);\\
S. G. J. Mochrie, M. Sutton, R. J. Birgeneou, D. E. Moncton and P. M. Horn, Phys. Rev. B {\bf 30},263 (1984).
\item[{[13]}] S. K. Stija, L. Passel, J. Eckart, W. Ellenson and H. Patterson, Phys. Rev. Lett. {\bf 51},411 (1983). 
\item[{[14]}] C. Ebner and W. F. Saam, Phys. Rev. A {\bf 22}, 2776 (1980);
     ibid, Phys. Rev. A {\bf 23},1925 (1981);
     ibid, Phys. Rev. B {\bf 28},2890 (1983).
\item[{[15]}] D. A. Huse , Phys. Rev. B {\bf 30},1371 (1984).

\item[{[16]}]  A. Benyoussef and H. Ez-Zahraouy, Physica A, {\bf 206}, 196
(1994).
\item[{[17]}]  A. Benyoussef and H. Ez-Zahraouy, J. Phys. {\it I } France 
{\bf 4}, 393 (1994).

\item[{[18]}] A. J. Freeman, J. Magn. Magn. Mater, {\bf 15-18}, 1070 (1980).

\item[{[19]}] R. Richter, J. G. Gay and J. R. Smith, Phys. Rev. Lett. {\bf 54}, 2704 (1985).
\item[{[20]}] R. H. Victora and L. M. Falicov, Phys. Rev. B {\bf 31}, 7335 (1985).
\item[{[21]}] C. Rau, C. Schneider, G. Xiang and K. Jamison, Phys. Rev. Lett. {\bf 57}, 3221 (1986).
\item[{[22]}] D. Pescia, G. Zampieri, G. L. Bona, R. F. Willis and F. Meier, Phys. Rev. Lett. {\bf 58}, 9 (1987).
\item[{[23]}] C. Rau, G. Xiang and C. Liu, Phys. Rev. Lett. A {\bf 135}, 227 (1989).
\item[{[24]}] L. Bahmad, A. Benyoussef and H. Ez-Zahraouy, Phys. Rev. E {\bf 66}, 056117 (2002).
\item[{[25]}] L. Bahmad, A. Benyoussef and H. Ez-Zahraouy,
M. J. Condensed Matter  {\bf 4}, 84 (2001).
\item[{[26]}] Q. Hong, Phys. Rev. B {\bf 41}, 9621 (1990); ibid Phys. Rev. B {\bf 46}, 3207 (1992).
\item[{[27]}] K. Binder and D.P. Landau, Phys. Rev. B {\bf 46}, 4844 (1992)
\item[{[28]}] K. Binder in {\it Cohesion and structure of Surfaces } vol. 4, Elsevier (1995).
\item[{[29]}] L. Bahmad, A. Benyoussef A. Boubekri and H. Ez-Zahraouy, Phys. Stat. Sol. b {\bf 215}, 1091 (1999).
 
\end{enumerate}

\noindent{\bf Figure Captions}\\
\\
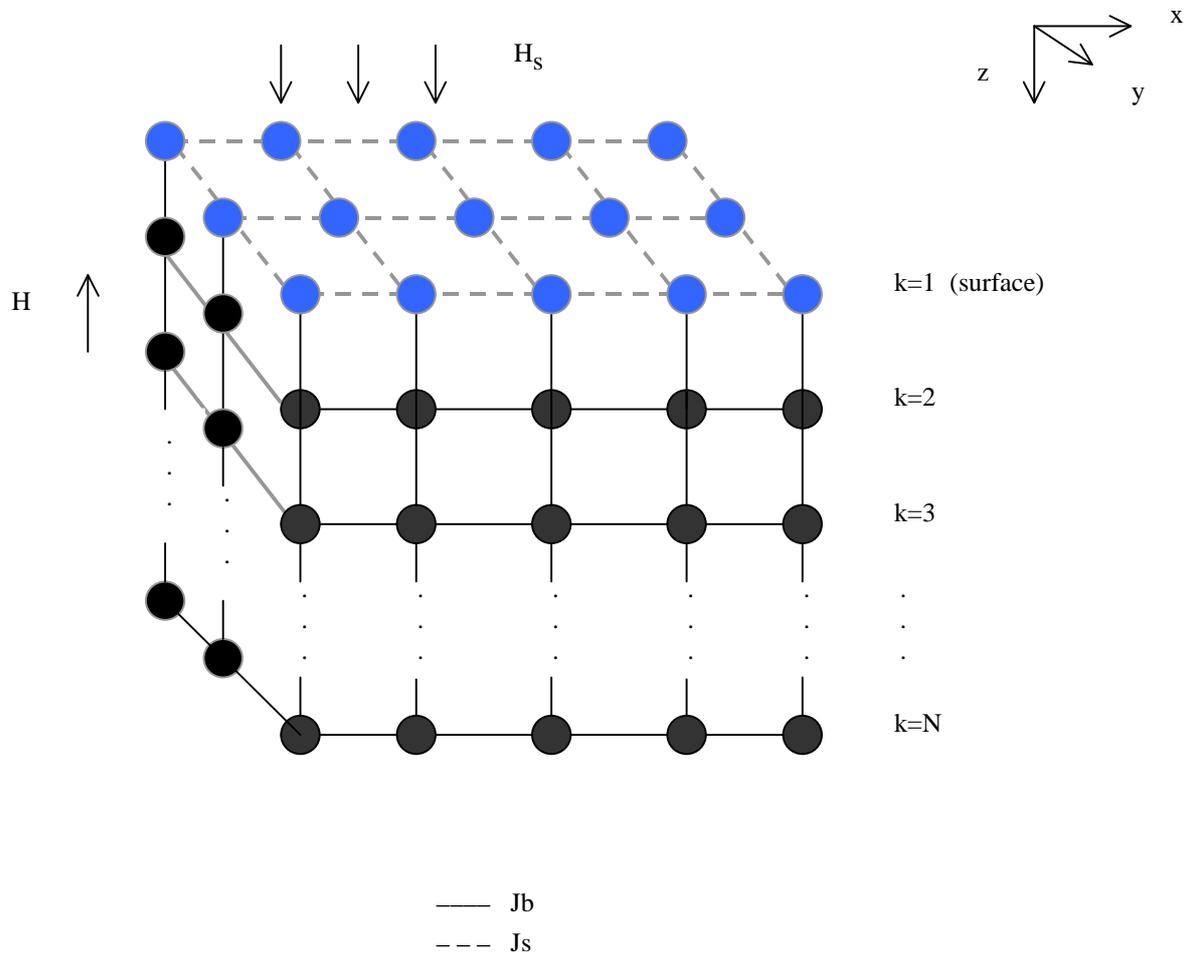
\noindent{\bf Figure 1.}
 A sketch of the geometry of the system formed with $N$ layers and subject to the surface magnetic field $H_s$ and the global magnetic field $H$. $J_s$ denotes the interaction coupling constant between the spins of the surface; $J_b$ is the interaction coupling constant between the spins of the bulk as well as between the spins of the bulk and those of the surface. \\

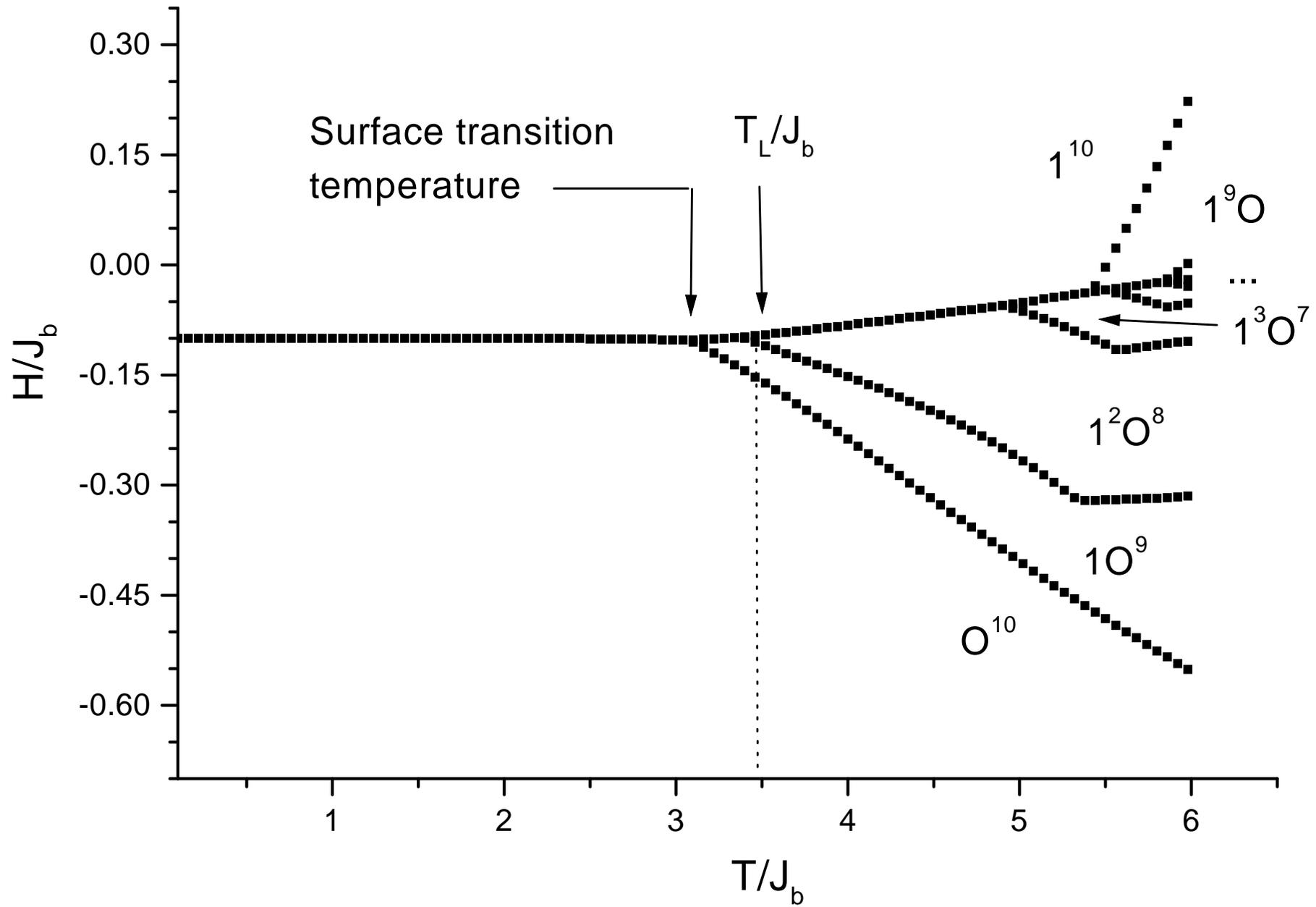
\noindent{\bf Figure 2.}: Phase diagram, in the $(H/J_b,T/J_b)$ plane, showing the layering transitions, and the definition of the layering transition temperature $T_L/J_b$, which is different from the surface transition temperature. This figure is plotted for a system size with $N=10$ layers, a surface magnetic field $H_s/J_b=1.0$ and a surface coupling constant $J_s/J_b=1.5$. \\

\noindent{\bf Figure 3.}: Layering transition temperature, $T_L/J_b$, behaviour as a function of the system size ( number of layers) with a) $H_s/J_b=0.1$ and b) $H_s/J_b=1.0$. The number accompanying each curve denotes the surface coupling $J_s/J_b$ value. \\

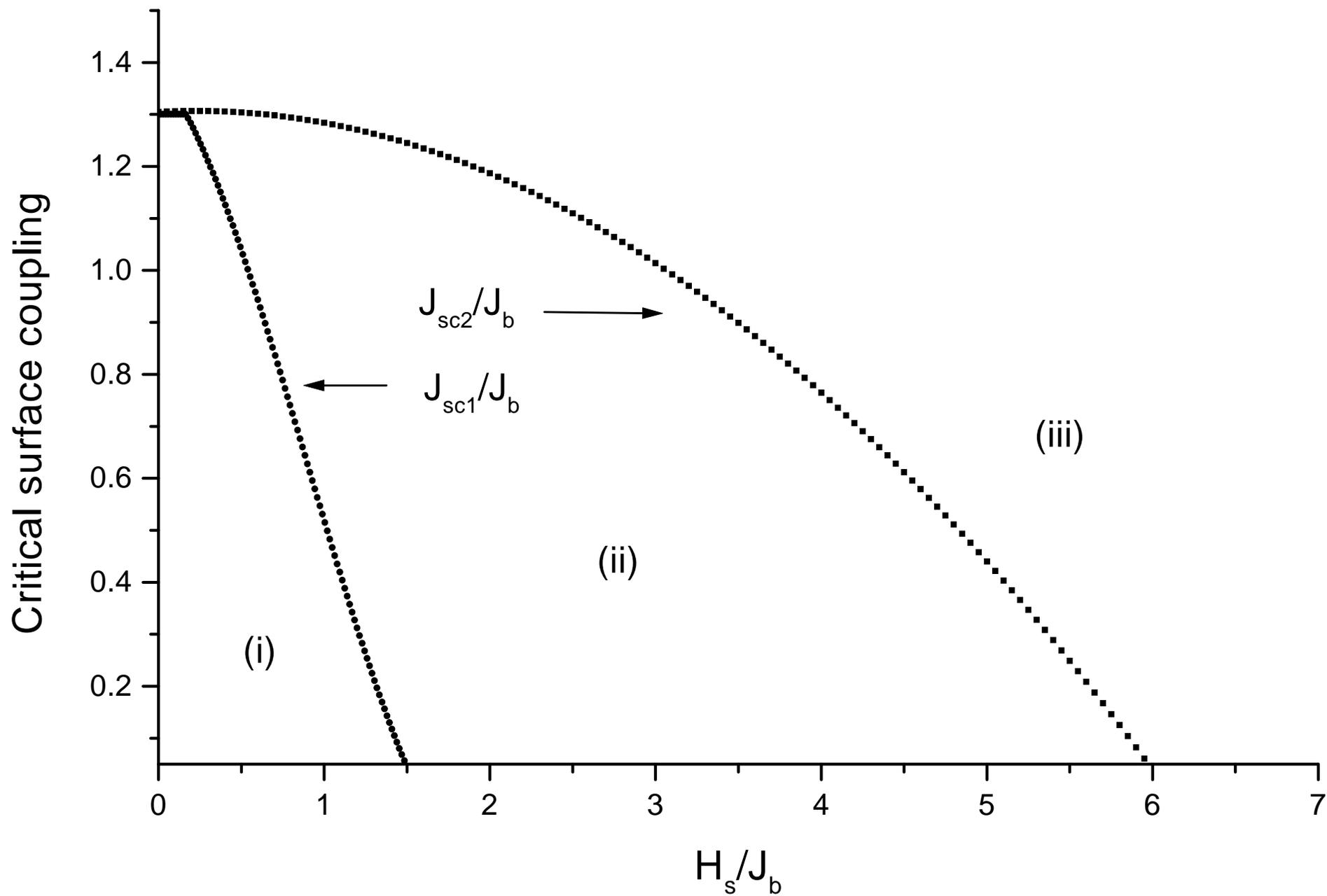
\noindent{\bf Figure 4.}: Critical surface couplings $J_{sc1}/J_b$ and $J_{sc2}/J_b$ profiles as a function of the surface magnetic field $H_s/J_b$. The behaviours $(i)$ and $(ii)$ are separated by the $J_{sc1}/J_b$ line; while $J_{sc2}/J_b$ separates the behaviours $(ii)$ and $(iii)$. \\

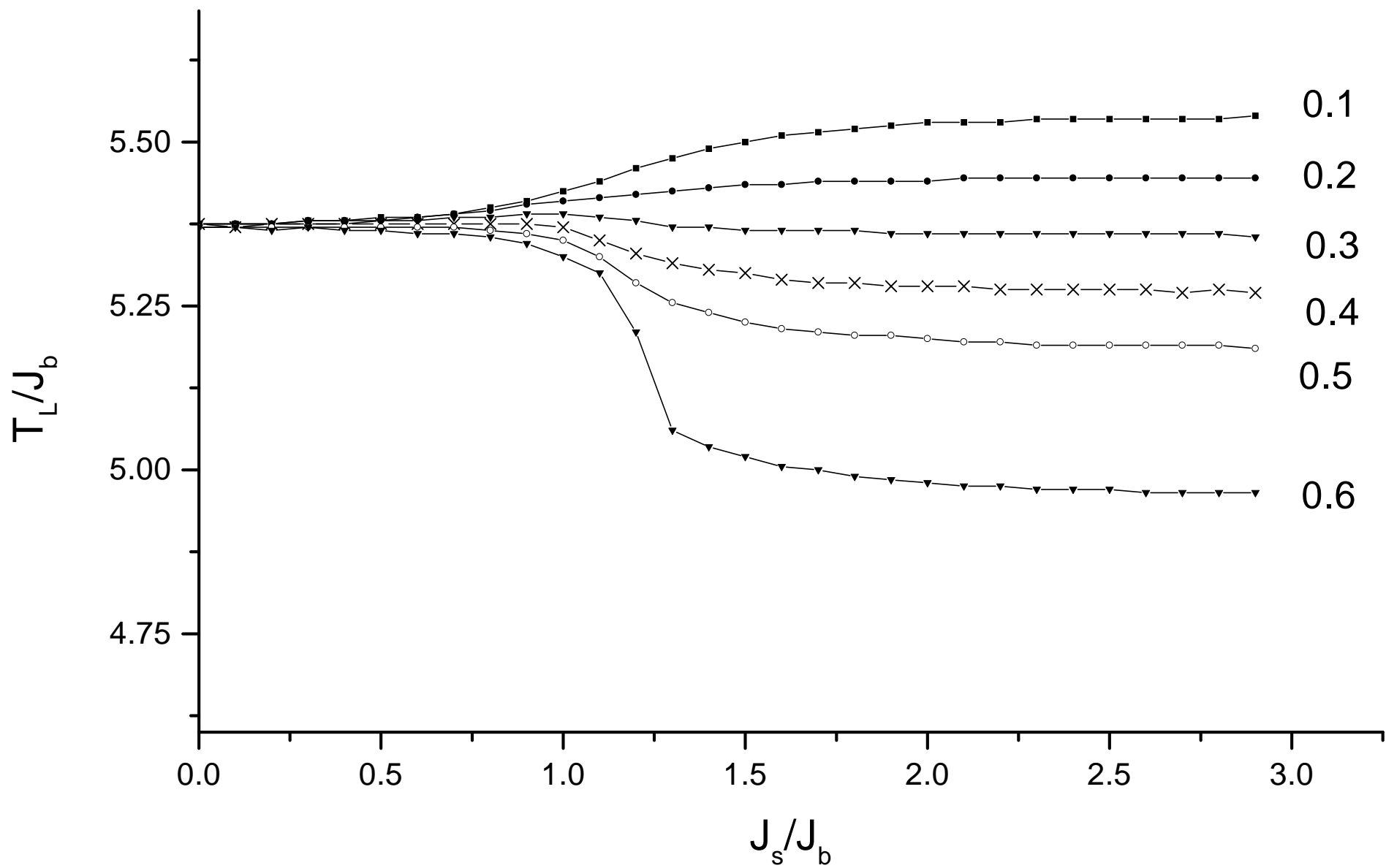
\noindent{\bf Figure 5.}: Layering transition temperature behaviour as a function of the surface coupling $J_s/J_b$ for $N=5$ layers and selected values of the surface magnetic field $H_s/J_b$: $0.1$, $=0.2$, $0.3$, $=0.4$, $0.5$ and $0.6$.

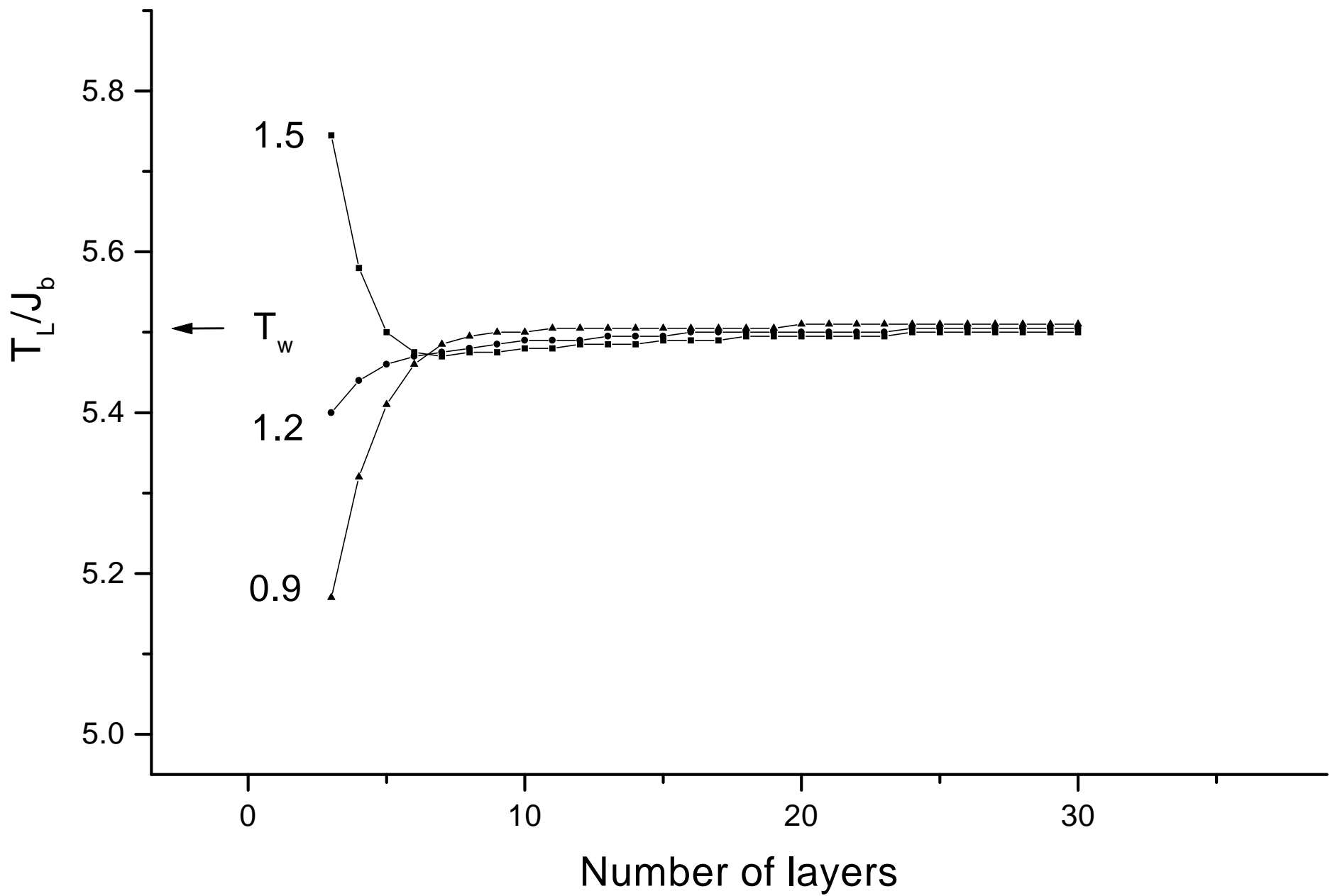
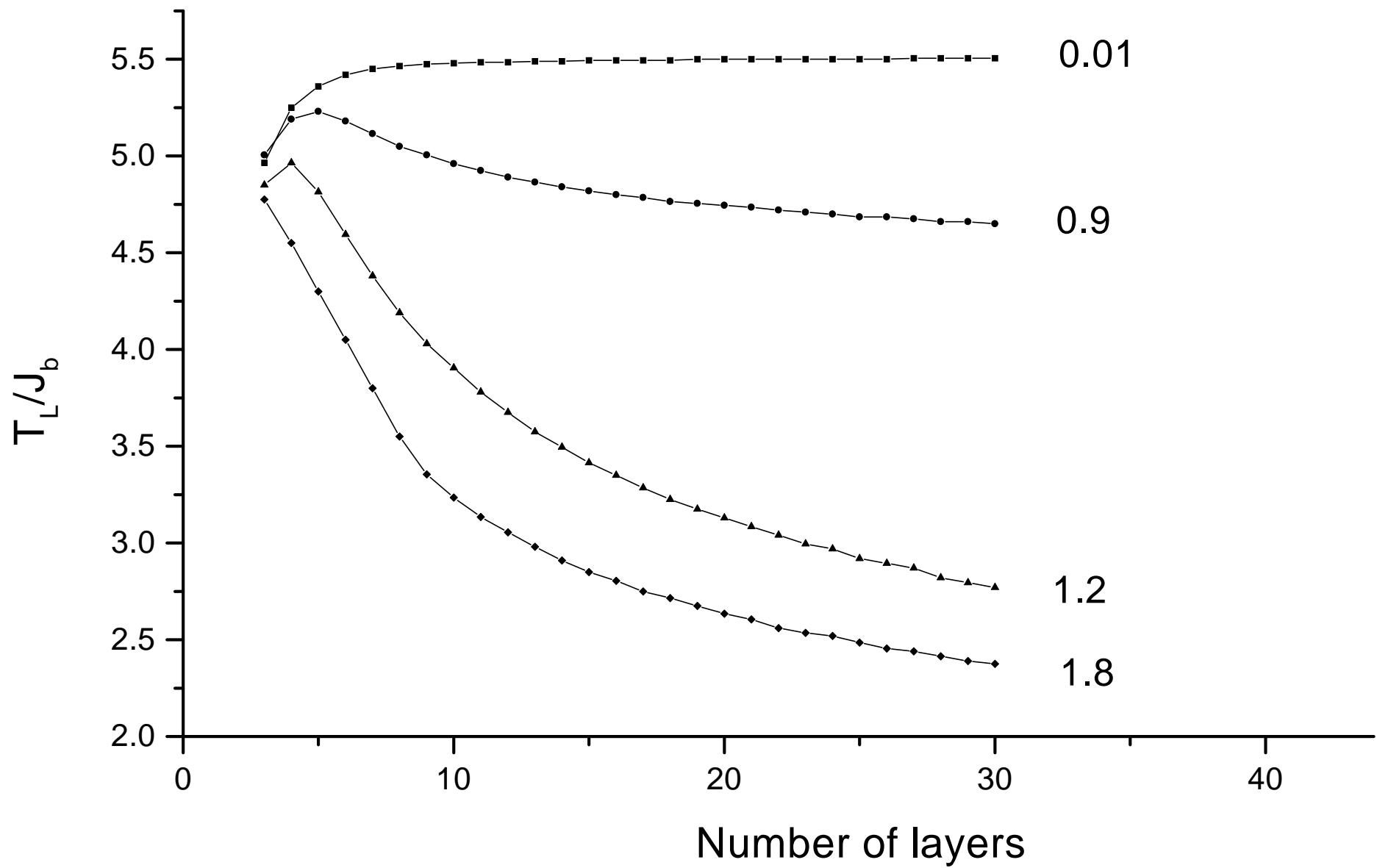
\end{document}